# COB-2023-0815
# NUMERICAL STUDY OF DISTORTED TULIP FLAME PROPAGATION IN CONFINED SYSTEMS

**Fernando Illacanchi, Sebastián Valencia, Cesar Celis**
Mechanical Engineering Section, Pontificia Universidad Católica del Perú
Av. Universitaria 1801, San Miguel, 15088, Lima, Peru
f.illacanchi@pucp.edu.pe, svalenciar@pucp.edu.pe, ccelis@pucp.edu.pe

**Armando Mendiburu Zevallos**
International Research Group for Energy Sustainability (IRGES)
Universidade Federal do Rio Grande do Sul
Rua Sarmento Leite 425, Porto Alegre, RS, Brazil
andresmendiburu@ufrgs.br

**Luis Bravo**
DEVCOM - US Army Research Laboratory
Aberdeen Proving Ground, MD 21005
luis.g.bravorobles.civ@army.mil

**Prashant Khare**
Department of Aerospace Engineering
University of Cincinnati
Cincinnati, OH 45221-0070
Prashant.Khare@uc.edu

*Abstract. Understanding the dynamics of premixed flames that propagates in confined systems is important in a wide range of applications. The study of premixed flames propagating in a closed channel covers a variety of thermochemical complexities related to flame ignition, laminar flame development, and strong non-linear interaction between the flame and the surrounding walls. Accordingly, to study the dynamics of premixed flames propagating in closed channels, numerical simulations of the propagation of distorted tulip flames are carried out in this work. All the numerical simulations are performed using the open-source computational tool PeleC, which is part of the Exascale Computing Project (ECP). More specifically, the fully reactive compressible Navier – Stokes equations are solved here using the high-order PPM (piecewise parabolic method). A 21-step chemical kinetic mechanism is employed to model the chemical kinetics and the energy release in a stoichiometric air/hydrogen mixture. Computational mesh independence studies are carried out in this work by both refining grid elements and employing different levels of adaptive mesh refinements (AMR). The final mesh employed here features an element size of 1/96 cm with 5 levels of refinement performed based on density gradients. The main results show that the classic tulip flame behavior evolves into a distorted one. Indeed, two consecutive collapses on the flame front are observed, which are related to wave pressure and the presence of reverse flow. Important aspects of the flame formation and propagation process analyzed include (i) the initial evolution of the tulip flame and its comparison with previous experimental and analytical results, (ii) the propagation of acoustic waves and its influence on flame evolution, and (iii) the formation of the distorted tulip flame and the collapse of flame cups. It is particularly found that the pressure wave produced by the interaction of the flame skirt with the side walls reduces the flame velocity and contributes to the formation of tulip flames. This is consistent with the reduction in both flame area and pressure gradient at the flame tip. Furthermore, the collapse of flame cups is associated with the vortex's formation near the channel side walls and the increase of pressure waves.*

## 1. INTRODUCTION

Understanding the dynamics of premixed flames and its relationship with more complex phenomena such as deflagration to detonation transition (DDT) is essential for a variety of engineering applications including pressure gain systems, such as rotating detonation engines (RDEs) (Dos et al., n.d.; Xiao et al., 2011). Due to the steep cost and limitations in experimental measurements, numerical simulations today provide a viable alternative that allow an investigation of the underpinning thermochemical mechanism. Flame propagation in closed and half-open channels is a complicated dynamic process because many physical phenomena play a key role during the flame ignition, the development of the laminar flame, and the interaction of the flame with the lateral walls reflecting pressure waves (Xiao et al., 2015). For instance, a laminar flame propagation is dominated by intrinsic instabilities such as the Darrieus-Landau (DL), thermal-diffusive, and Rayleigh-Taylor (RT) ones (Chung, 2006; Xiao et al., 2015).

Several past studies about premixed flames propagating in closed tubes discussed a series of flame shape changes such as spherical, finger, concave, and convex shapes. During the associated flame propagation processes, two main flame configurations are usually identified, (i) tulip and (ii) distorted tulip flames. The first one is a concave flame front with cusps at the lateral walls pointing toward the unburned mixture (Xiao et al., 2017a). The first photograph of this flame was obtained by Ellis in 1928 (Ellis, 1928) and it was first named as tulip flame by Salamandra in 1959, who noticed the required flame aspect ratio to successfully reproduce the flame inversion (length/diameter > 2 for closed tubes) (Clanet & Searby, 1996). The flame transition to a tulip flame is accompanied by a flame surface area reduction and a velocity decrease. Over the years different physical mechanisms have been proposed to explain the flame front inversion from a convex shape to a concave one. For instance, based on an experimental work, Ponizy et al. (2014) proposed that the tulip flame is purely governed by a hydrodynamic process without any dependency on intrinsic instabilities. In contrast, after carrying out a theoretical and experimental study, Clanet and Searby (1996) suggested that the flame inversion is governed by Taylor's instabilities at the flame skirt right after the flame front inversion.

The second flame configuration, distorted tulip flame (DTF), was observed more recently by Xiao et al. (2011) in a hydrogen/air mixture. In the referred work it is concluded that this flame can be observed in both closed and half-open channels, but in the latter ones, DTF is reproducible only in equivalence ratios in the range of $0.84 \leq \varphi \leq 4.22$. Notice that a DTF forms from consecutive collapses of the cusps near the channel lateral walls right after a well-noticeable tulip flame. The formation of this flame is usually associated with intrinsic instabilities such as RT ones and pressure waves generated from the interaction between the flame skirt and the duct side walls (Xiao et al., 2015). Nevertheless, various complex physical phenomena are involved in the formation of a DTF such as those coming from viscosity effects and other intrinsic instabilities.

Accordingly, in the present work, a numerical study of propagating flames in closed tubes is carried out. More specifically, using a 21-step chemical mechanism, this numerical study aims to elucidate the dynamics of flame formation and propagation, especially that prevailing in tulip and distorted tulip flames. Likewise, this study assesses the capabilities of the computational tool PeleC to properly describe the reactive flows accounted for here. Summarizing what follows, sections 2 and 3 describe, respectively, the mathematical and numerical models employed here. And sections 4 and 5 the main results obtained in this work and the conclusions drawn from them, respectively.

## 2. MATHEMATICAL MODELING

The numerical simulations carried out here involved the solution of the 2D fully compressible reactive Navier-Stokes equations, coupled with a 21-step kinetic mechanism developed by Li et al. (2004) to describe the chemical kinetics. Notice that this kinetic mechanism is suited for nitrogen as the bath gas. Accordingly, the governing equations used in this work to describe the reacting flow (Henry de Frahan et al., 2022; Marc T & Jon S Rood, n.d.; Poinsot & Veynante, n.d.) read as follows,

$$\frac{\partial \rho}{\partial t} + \frac{\partial \rho u_i}{\partial x_i} = 0 \tag{1}$$

$$\frac{\partial \rho Y_k}{\partial t} + \frac{\partial (\rho(u_i + V_{k,i})Y_k)}{\partial x_i} = \dot{\omega}_k \tag{2}$$

$$\frac{\partial \rho u_j}{\partial t} + \frac{\partial p}{\partial x_j} + \frac{\partial (\rho u_i u_j)}{\partial x_i} = \frac{\partial T_{ij}}{\partial x_i} + \rho \sum_{k=1}^{N} Y_k f_{k,j} \tag{3}$$

$$\frac{\partial (\rho E)}{\partial t} + \frac{\partial (\rho E u_i + p u_i)}{\partial x_i} = \frac{\partial (u_i T_{ij})}{\partial x_j} - \frac{\partial q_i}{\partial x_i} + \dot{\omega}_T + \rho \sum_{k=1}^{N} Y_k f_{k,i}(u_i + V_{k,i}) \tag{4}$$

$$\dot{\omega}_T = \sum_{k=1}^{N} \Delta h_{f,k} \dot{\omega}_k \tag{5}$$

$$q_i = -k\frac{\partial T}{\partial x_i} + \rho \sum_{k=1}^{N} h_k Y_k V_{k,i} \tag{6}$$

$$\tau_{ij} = \frac{-2}{3}\mu \frac{\partial u_k}{\partial x_k}\delta_{ij} + \mu\left(\frac{\partial u_i}{\partial x_j} + \frac{\partial u_j}{\partial x_i}\right) \tag{7}$$

$$E = \sum_{k=1}^{N} Y_k E_k(T) \tag{8}$$

$$p = RT \sum_{k=1}^{N} \frac{Y_k}{W_k} \frac{1}{\tau - b_m} - \frac{a_m}{\tau(\tau + b_m)} \qquad (9)$$

where $\rho, u_i, Y_k, t, \dot{\omega}_k, p, E, \dot{\omega}_T, R, W_k$ and $T$ stand for density, velocity, mass fraction, time, reaction rate of species $k$, pressure, total energy, heat release by combustion, universal gas constant, molecular weight of species $k$, and temperature, respectively. The diffusive velocity along the $i$ direction of species $k$, $V_{k,i}$, is included in the mass conservation equation for species $k$, Eq. (2). The viscous stress tensor, $\tau_{ij}$, is defined in turn in Eq. (7), where $\mu$ is the dynamic viscosity. In the linear momentum equation, Eq. (3), the body force acting on species $k$ along the $i$ direction is also considered. Notice as well that the energy conservation equation, Eq. (4), includes the power produced by the body force acting on species $k$, $\rho \sum_{k=1}^{N} Y_k f_{k,i}(u_i + V_{k,i})$. In addition, the energy-diffusive term $q_i$, Eq. (6), comprehend the diffusion term given by Fourier's law, and a second term associated with diffusion of species with different enthalpy. To close the system involving the compressible Navier-Stokes equations, the gas equation of state is employed. In this work, the Soave-Redlich-Kwong (SRK) equation of state, Eq. (9), has been utilized, where $a_m$ and $b_m$ represent the repulsion terms.

Transport properties, including dynamic viscosity, mass diffusivity, and thermal diffusivity, are calculated in turn using the relationships developed by Ern and Givangigli (1995),

$$\mu = \sum_{m=1}^{N} (X_m(\mu_m)^\alpha)^{\frac{1}{\alpha}} \quad with \alpha = 6 \qquad (10)$$

$$k = \sum_{m=1}^{N} (X_m(k_m)^\alpha)^{\frac{1}{\alpha}} \quad with \alpha = 1/4 \qquad (11)$$

$$D_{m,mix} = \frac{1 - Y_m}{\sum_{j \neq m} \frac{X_j}{D_{mj}}} \qquad (12)$$

Finally, as indicated above, the combustion of the stoichiometric hydrogen/air mixtures considered here is described using a 21-step chemical kinetic mechanism, which includes 8 chemical species, $H_2, O_2, H_2O, H, O, OH, HO_2, H_2O_2$.

## 3. NUMERICAL APPROACH

The numerical simulations are carried out here using the open-source computational tool PeleC (Marc T & Jon S Rood, n.d.), which provides a platform for combustion and turbulence-chemistry interaction research in the ExaScale computing era (Marc T & Jon S Rood, n.d.). Since it allows focusing the computational resources on specific flow regions, a block-structured adaptative mesh refinement (AMR) method is also employed. Notice that the AMR processes performed do not obey a fixed relationship between coarser and finer grids. That is, the refinements vary over time as a function of a flag or tracer such as the temperature gradient (Zhang et al., 2019). In terms of numerical schemes, the PPM (piecewise parabolic method) one extensively described in (Colella & Woodward, 1984) is used in this work. This scheme, an extension of Godunov's method first introduced by van Leer (van Leer, 1979), is well suited for computing strong shocks.

### 3.1. Computational domain and boundary conditions

A closed tube, 4 cm wide and 28 cm long, is numerical studied in this work. The corresponding 2D computational domain is shown in Figure 1. More specifically, following previous works (Li et al., 2021), a half tube with a symmetric axis is considered here. Therefore, the upper boundary condition ($y = 2$ cm) corresponds to a symmetry condition. In addition, to focus exclusively on the flame propagation dynamics, all the walls are considered adiabatic, non-slip, and reflecting. The ignition is performed in turn using a semi-circular pocket of hot gas at a specified temperature, $T_{ignition}$. The radius of this hot gas pocket is determined from the theoretical study by Bychkov et al. (2007), where a dimensionless analysis for the early stages of the flame propagation process is proposed.

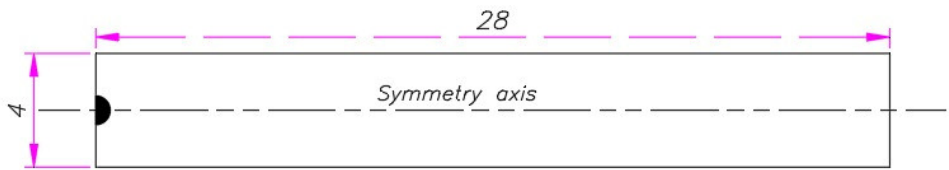

**Figure 1 Closed tube computational domain. All walls are adiabatic, non-slip, and reflecting. Dimensions in cm.**

**3.2. Mesh refinement criteria**

The PeleC solver (Marc T & Jon S Rood, n.d.) includes an adaptive mesh refinement (AMR) method, which allows the dynamic creation of grid levels based on various tagging criteria such as species mass fractions or density gradient (Marc T & Jon S Rood, n.d.; Zhang et al., 2019). The associated grid refinement is performed according to criteria defined by the user, so the solver allows limitless levels of refinement. In this work, two tagging criteria, 5 levels of refinement, and a refinement ratio of 8 for each level, have been employed. Following verification examples of the solver (Marc T & Jon S Rood, n.d.), $HO_2$ has been used as the chemical species that characterizes the reaction zone. Thus, mesh refinements have been performed where the mass fraction of $HO_2$ is above a specified value. Similarly, due to the density jump at the flame front, additional mesh refinement criteria based on the density ratio increase, allowing an optimization of the computational cost and a reduction of the numerical errors coming from the initial discretization of the flame front (Figure 2), have been also utilized. These criteria are expressed as follows,

$$Max\left(\left\|\frac{f_{i+1,j}}{f_{i,j}}\right\|, \left\|\frac{f_{i,j+1}}{f_{i,j}}\right\|, \left\|\frac{f_{i,j}}{f_{i-1,j}}\right\|, \left\|\frac{f_{i,j}}{f_{i,j-1}}\right\|\right) \geq u \tag{13}$$

$$f_{i,j} \geq u \tag{14}$$

where $f_{i,j}$ is a computed parameter at a given control volume $i, j$. Eq. (13) is applied for the density jump at the flame front and the configured ratio is 1.02. Finally, the $HO_2$ threshold accounted for is equal 2.5e-4.

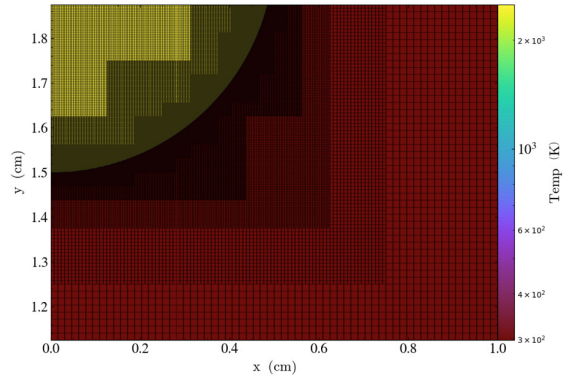

**Figure 2** Details of the computational mesh at the start of the numerical simulations.

**4. RESULTS AND DISCUSSIONS**

**4.1 Chemical kinetic mechanism**

To reproduce major characteristics of air/hydrogen mixtures, previous works involving flame propagation in closed ducts mostly employed ideal one-dimensional steady-state flame models with one-step global reactions (Gamezo et al., 2007; Xiao et al., 2015). This type of models, widely validated experimentally in (Oran & Gamezo, 2007) for instance, have been mainly employed in large eddy simulation (LES) contexts (Han et al., 2019) to reduce the associated computational cost. In the present work, finite rate chemistry is accounted for through the use of a 21-step chemical kinetic mechanism for nitrogen as a main dilutant (Marc T & Jon S Rood, n.d.). Accordingly, Figure 3 shows the temporal evolution of the ratio between the temperature and density of the unburned mixture and the burnt one. Notice that the variation in time of both temperature and density ratio, whose initial values agree with the input model parameters defined in (Gamezo et al., 2007; Xiao et al., 2015), comes from the compressibility effects caused by the overall pressure and temperature increase in the closed channel. Additionally, the temporal variation of the flame thickness at the leading flame-tip is shown in Figure 4. The flame thickness is computed by the temperature profile, $\delta_L^o = \frac{T_2 - T_1}{\max\left(\frac{\partial T}{\partial x}\right)}$, (Poinsot & Veynante, n.d.). The rapid decrease of this parameter is due to the behavior of the density and temperature ratios, as well as to the increase of the overall pressure over time, also shown in Figure 4. These results emphasize that the employed chemical kinetic mechanism properly describes the main characteristics of the air/hydrogen reacting mixture studied here.

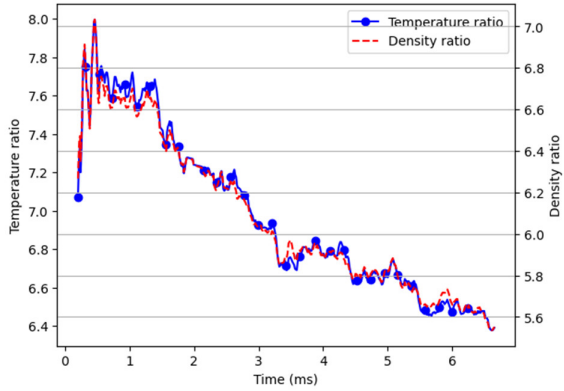
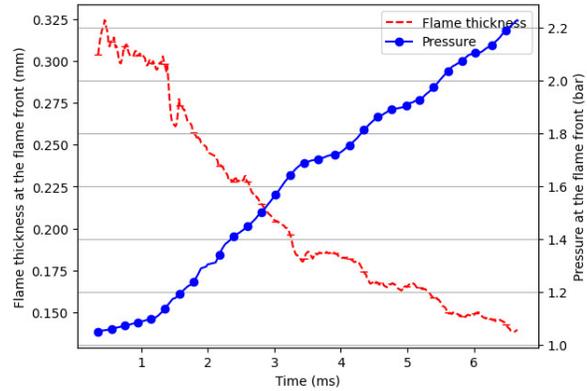

Figure 3 Temporal evolution of temperature and density ratio (defined as unburned mixture/burnt mixture).

Figure 4 Temporal evolution of flame thickness at the leading flame-tip and pressure at the right-end wall.

**4.2. Analytical and numerical results of flame propagation in early stages**

The acceleration mechanism of premixed flames in open channels was previously studied experimentally by Clanet and Searby (1996). Similarly, Bychkov et al. (2007) proposed an analytical model for the acceleration of premixed flames at the early stages accounting for cylindrical channels with slip adiabatic walls. This analytical model employs dimensionless coordinates for position $\frac{(r,z)}{R}$, velocity $\frac{(u_r,u_z)}{u_f}$, and time $\tau = \frac{U_f t}{R}$. In addition, it considers an uncompressible mixture that ignites at the center of the closed wall and propagates towards the open end. In particular, the referred analytical model describes accurately the initial stages of a propagating premixed flame, that is, the initial hot chunk of gas that evolves into a finger shape flame until the skirt's flame touches the lateral walls. In their model, Bychkov et al. (2007) established reduced times for each main stage of the flame propagation process. Accordingly, the spherical shape is defined when the flame skirt moves halfway to the side wall $\tau_{sph} = \frac{1}{2\alpha}$, where $\alpha = \sqrt{\Theta(\Theta-1)}$ and $\Theta$ is the expansion factor $\left(\frac{\rho_u}{\rho_b}\right)$, defined as the density ratio of the unburned and burnt gas (Bychkov et al., 2007). The reduced time at which the flame skirt touches the lateral wall is in turn defined as,

$$\tau_{wall} = \frac{1}{2\alpha}\left(\frac{\Theta + \alpha}{\Theta - \alpha}\right). \tag{15}$$

Later Xiao et al. (Xiao et al., 2012) suggested a scaled time for the tulip flame phenomenon,

$$\tau_{tulip} = \tau_{inv} + \tau_{wall}, \tag{16}$$

where $\tau_{inv} = \lambda/\alpha$ is the period between the $\tau_{wall}$ and $\tau_{tulip}$. Finally, the position of the leading flame-tip is defined as (Bychkov et al., 2007),

$$Z_{tip} = \frac{\Theta}{4\alpha}[\exp(2\alpha\tau) - \exp(-2\alpha\tau)] = \frac{\Theta}{2\alpha}\sinh(2\alpha\tau). \tag{17}$$

This theoretical model has been validated in the past using the experimental data obtained by Clanet and Searby (1996) and the agreement has been relatively good (Bychkov et al., 2007). Figure 5 and Figure 6 show, respectively, a comparison of the reduced leading flame-tip position and displacement velocity between the described theoretical model and the numerical results obtained using PeleC. As noticed from these figures, initially the analytical and numerical results show a good agreement. However, since the theoretical model considers a constant pressure, non-compressible phenomena, and constant burning velocity (Xiao et al., 2012), it tends to overestimate the velocity, consequently, the position of the leading flame-tip (Figure 5). Thus, the analytical model is only valid for the early stages, where intrinsic instabilities, pressure waves, and turbulence do not play a key role in the flame propagation process. From Figure 6, due to the interaction of the flame front with the reflecting pressure wave originated at the right-end wall, at a reduced time of $\tau = 0.147$, the displacement velocity computed numerically largely differs from the analytical one. This is an expected result because pressure wave interactions are not considered in the analytical model proposed by Bychov et al. (2007).

It is worth noticing here that the combustion process studied in this work generates weak pressure waves that propagate out ahead of the flame front, which are reflected by the tube lateral walls. However, these initial waves are not strong enough to greatly alter the flame propagation dynamics (Xiao et al., 2015). The first reflected pressure waves from the right end wall travels back through the channel and interact with the flame front at $1.28\ ms$, as shown in Figure 7. This reflecting wave, unlike the initial ones coming from the lateral walls, reduces considerably the velocity displacement of the leading tip-flame, from 30 m/s to 15 m/s (Figure 6). Once the reflection reaches the flame front, the pressure wave passes through the reaction zone and strongly increases the local vorticity, mainly at the flame front. At the same time, the flame front generates a weak reflecting wave. Finally, the first reflecting wave is reflected by the left-end wall of the tube to the unburned mixture and, as shown in Figure 7 at 1.29 ms, the flame front reflects immediately a weak pressure wave.

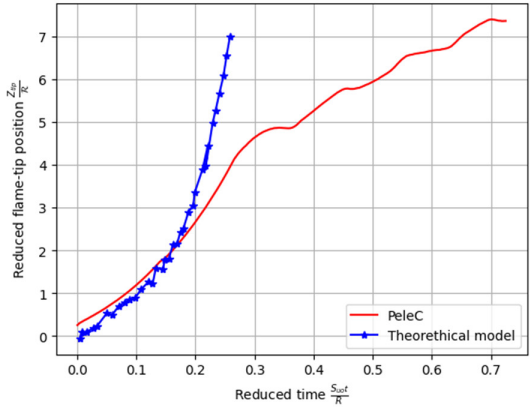

Figure 5 Reduced position of the leading flame tip computed theoretically and using PeleC.

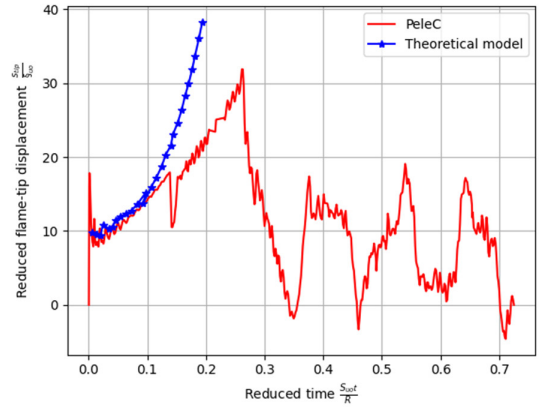

Figure 6 Reduced displacement velocity of the leading flame tip computed theoretically and using PeleC.

Figure 8 highlights the influence of intrinsic instabilities such as the Darrieus-Landau (DL) on flame dynamics (Clavin & Searby, 2016). In premixed propagating flames, the difference in density of the fresh and burnt gases is responsible for the piston effect that pushes away the fresh gas to the right-end wall. Figure 4 shows in particular that the expansion ratio varies according to the overall pressure. Moreover, there are some local pressure variations due to the interaction between the flame front and the reflecting waves. Additionally, there are nonlocal hydrodynamics and thermos-diffusive instabilities which are intrinsic of flame front. For instance, the flow streamlines are deflected by the curvature of the flame front, DL instabilities. Thus, as shown in Figure 8, the vorticity in the burnt gas is concentrated mainly at the back of the reaction zone where the curvature is the highest. It is worth noticing that the maximum value of vorticity is concentrated in the reaction zone, thermal-diffusive instabilities. This perturbation increases over time and generated wrinkles at the flame front, as is shown in Figure 10.

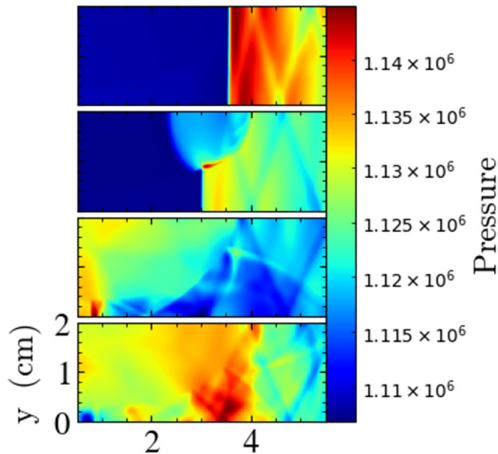

Figure 7 Pressure field and reflecting waves shocking the flame front at times (from top to bottom) t = 1.28, 1.29, 1.35, and 1.40 ms, respectively.

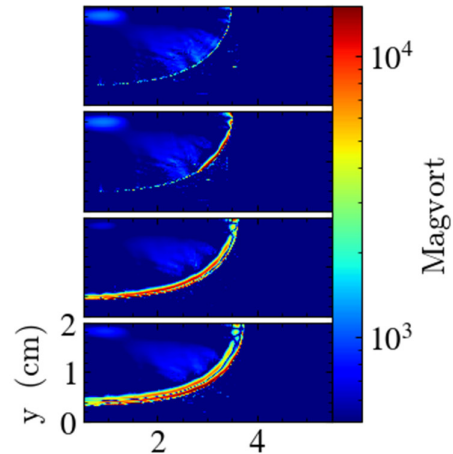

Figure 8 Vorticity distribution at times (from top to bottom) t = 1.28, 1.29, 1.35, and 1.40 ms, respectively.

### 4.3. Overall evolution and shape change of the flame

Figure 9 shows Schlieren photographs (left) and numerical predictions (right) of flame temperatures at representative times. The flame propagation is usually divided into 5 stages, (i) spherical flame, (ii) finger-shape flame, (iii) front flame skirt touching lateral walls, (iv) tulip flame (TF), and (v) distorted tulip flame (DTF). After the flame ignition, the flame front expands uniformly and, due to the combustion process, pressure waves that travel out ahead of the flame front are generated. These pressure waves reach the lateral walls and are reflected, creating a series of crisscrossed lines representing local pressure increases. More specifically, in the first stage of the flame propagation process, the flame front expands without any effects on the lateral walls. Then, at $t = 0.52\ ms$, when the flame front reaches halfway to the tube lateral wall (Bychkov et al., 2007), the transition from spherical to finger shape takes place. In the second stage, due to the enclosure of the lateral walls (Kurdyumov & Matalon, 2015), the flame front is elongated in the axial direction and the flame surface area increases considerably, consequently, so does the displacement velocity. The flame front first touches the tube side wall at $t = 2.34\ m$ This flame-wall interaction generates pressure waves that travel along the tube and interact with the flame front. The velocity displacement of the leading tip is consequently reduced considerably due to the generated pressure waves and the surface reduction. During this third stage, there is a constant pressure wave production by the interaction of the flame skirt with the lateral walls. As is shown in Figure 9, the flattened flame front is formed at $t = 3.3\ ms$ and, thereafter, the flame front inversion takes place. This last phenomenon occurs mainly due to both the reverse flow generated at the central region and the hydrodynamic effects of the lateral pressure waves. The tulip flame characterized by a well-pronounced cusp pointing to the right end wall is formed at $t = 4.13\ ms$. After the formation of the TF, secondary cusps are formed at the flame front near the lateral walls, which move towards the central axis. The Schlieren images show the referred motion inside the burnt gas after the formation of the flattened flame front ($t = 3.3\ ms$). In the central region, the flow travels in the opposite direction of the leading flame-tip and forms a mushroom structure due to compression of the lighter burnt gas.

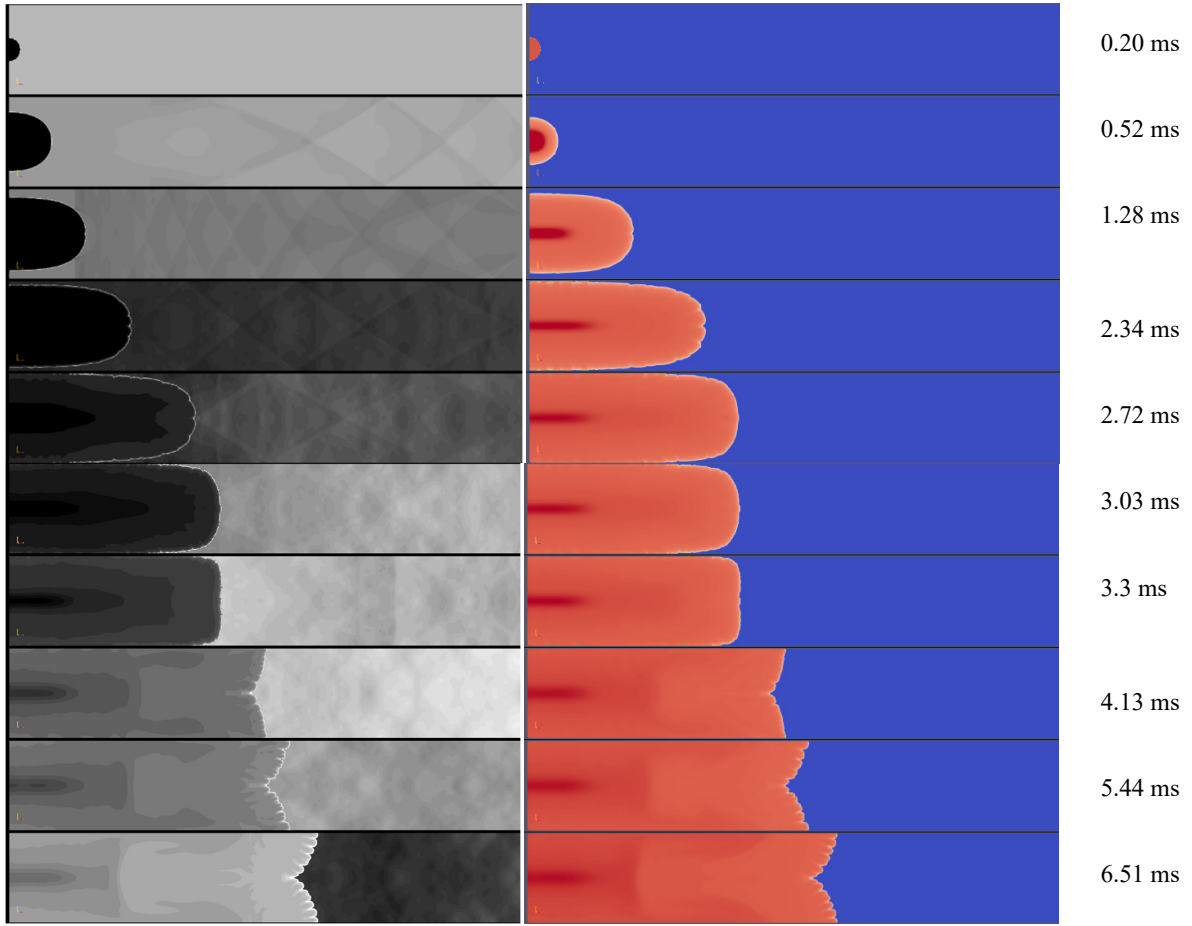

**Figure 9 Sequence of Schlieren photographs (left) and numerical predictions (right) of temperature fields showing the evolution of premixed air/hydrogen mixtures in a closed channel of 4 cm x 28 cm.**

### 4.4. Generation of pressure waves

Following previous works (Xiao et al., 2015, 2017a), the generation of pressure waves and their interaction with the flame front in premixed flames has also been analyzed in this work. Accordingly, Figure 10 shows temperature (left) and pressure (right) fields at the third stage of the flame propagation process, finger shape flame, at which the flame front touches the lateral walls and generates pressure waves. Indeed, after the flame skirt reaches the side wall, at $t = 2.34\ ms$, a semi-circular rarefaction wave that propagates front and back through the channel is generated. This rarefaction wave reaches the leading flame front at $t = 2.44\ ms$, which increases locally the pressure and reduces the displacement velocity. The contact between the flame skirt and the sides walls also generates a reverse flow in the burnt region (Figure 15). Additionally, the flame surface area reduces considerably the displacement velocity (Xiao et al., 2015). When the flame front elongates due to the enclosure of the lateral walls, local wrinkles along the surface flame front are formed. Specifically, at the time the flame skirt reaches the lateral wall, three wrinkles generate the rarefaction waves, Figure 10 ($t = 2.34\ ms$). The second interaction between the flame front and the lateral walls generates a stronger rarefaction, Figure 10 ($t = 2.72\ ms$), which, similar to the first rarefaction wave, decelerates the leading flame tip more intensively. Finally, there is a third flame-wall interaction that generates a weak wave. This is produced near the time of formation of the flattened flame and does not impact considerably the flame propagation dynamics.

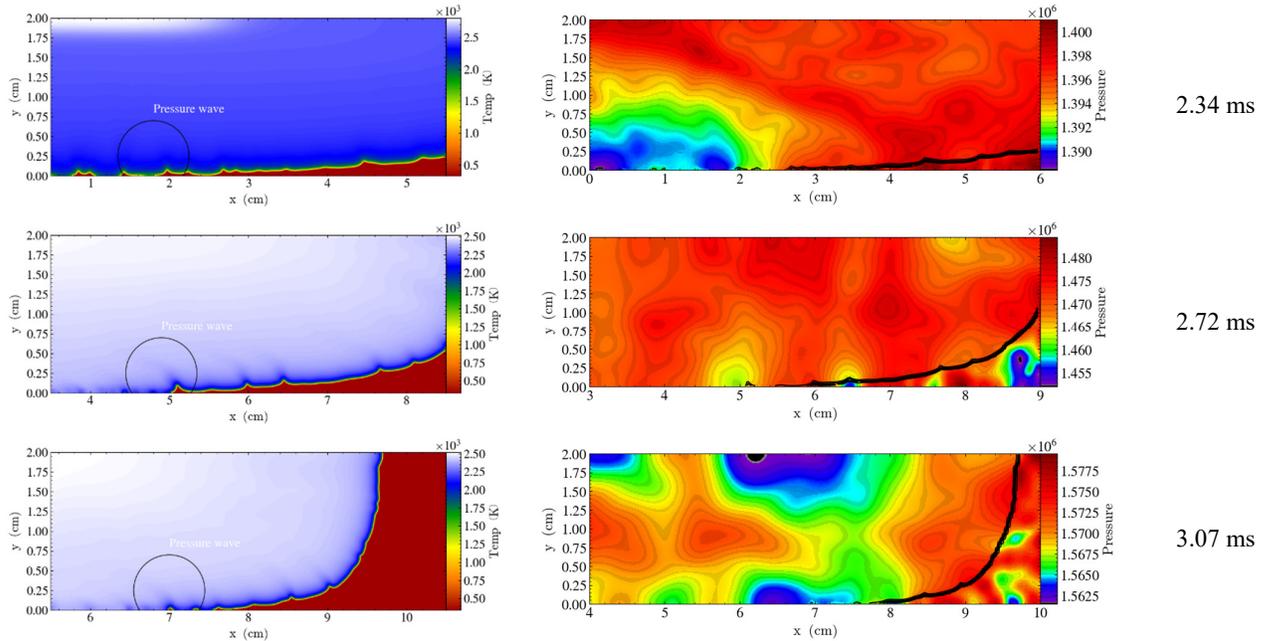

**Figure 10** Temperature (left) and pressure (right) contours, and pressure wave generation.

### 4.5. Flame propagation dynamics

In this section, the flame propagation dynamics is discussed in terms of the leading flame-tip, which is defined as the nearest point of the flame front to the right-end wall. Thus, unlike previous works (X. Li et al., 2021; Xiao et al., 2013, 2017b), where the leading tip moves initially along the channel centerline, the y-axis position of the leading tip in this work moves slightly away from the centerline. This occurs because of the interaction between the flame front and the pressure waves present in the flow (Figure 7), i.e., the displacement velocity is temporally reduced due to the pressure waves that shocks the leading flame-tip. Consequently, the position of the leading flame-tip moves down. After the flame inversion, the flame front near the lateral walls moves ahead of the one at the centerline, thus, the leading flame-tip moves near the sidewalls, where the cusps are formed. To illustrate this point, Figure 12 shows the temporal evolution of both the position of the leading flame-tip and the displacement velocity. Notice that the flame position here is defined as the maximum distance of the reaction zone to the left end wall, and the origin of the coordinate system is set at the center of the left end wall. The displacement velocity is defined thus as the variation of the flame position over time and computed by a centered difference. As shown in Figure 14, and in agreement with the results shown in (Xiao et al., 2013)

It is worth emphasizing here that the flame propagation velocity fluctuates greatly temporally during the combustion process. This local acceleration and deceleration lead to the formation of wrinkles at the flame front, and these perturbations are amplified along the flame development (Clavin & Searby, 2016; Xiao et al., 2013). However, the overall velocity profile is described by a well-defined curve that depicts the development of the flame. Initially, due to the flame front surface increase, the flame accelerates rapidly until the formation of the finger-shape flame. At $t = 1.28\ ms$, the first reflecting wave from the end wall reaches the flame front and strongly decreases the displacement velocity (Figure 14). In contrast, there is a temporal strong increase of the pressure growth. Unlike previous works (Xiao et al., 2012), this first reflecting wave plays a key role in the initial development of the flame since this wave retards the time when the flame skirt touches the lateral walls. The flame skirt touches the side walls at $t = 2.34\ ms$ and this flame-wall interaction generates a rarefaction wave that reaches the flame front. These waves and the surface area reduction govern the dynamics of the flattened flame. A second rarefaction wave is generated at $t = 2.72\ ms$ and this reinforces and extends the flame deceleration until $t = 3.03\ ms$. At this time, the flattened flame is noticeably visible (Figure 9) and starts the inversion of the flame, that is, the flame front near the side walls moves ahead of the flame at the centerline. According to Markstein (1953), flame inversion is due to the deceleration of the flame front and the interaction of pressure waves. In the present work, no strong pressure waves are acting in the closed tube, however, the pressure waves generated during the development of the TF are comparable.

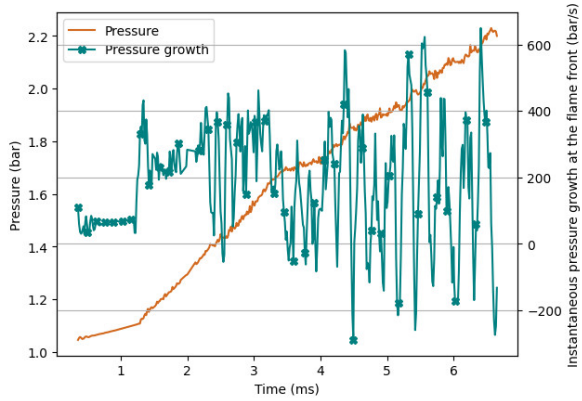

**Figure 11 Pressure and pressure growth at the leading flame-tip over time.**

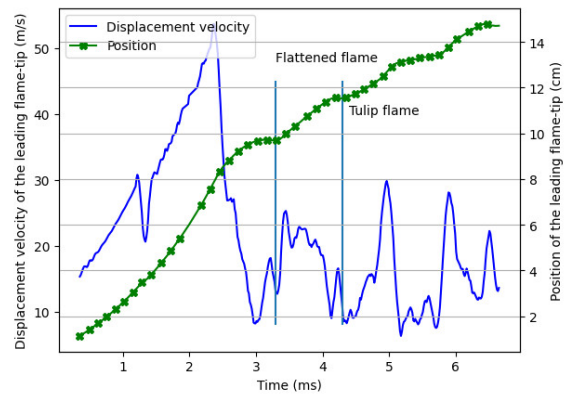

**Figure 12 Displacement velocity and position of the leading flame-tip over time.**

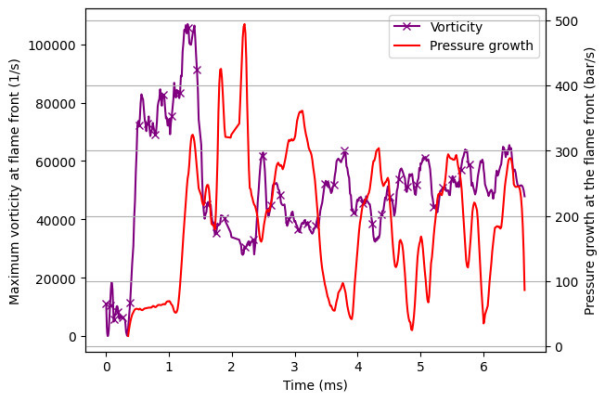

**Figure 13 Maximum vorticity at the flame front and pressure growth at the leading flame-tip over time.**

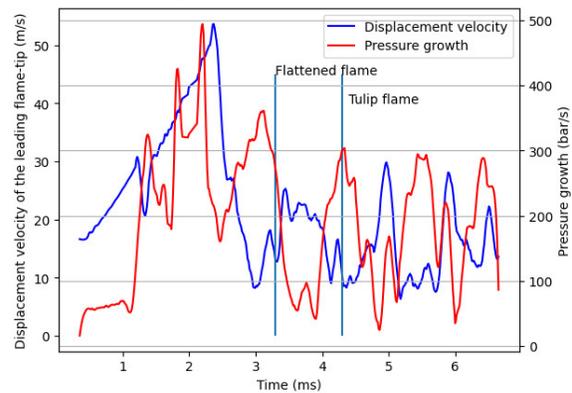

**Figure 14 Displacement velocity of the leading flame-tip and pressure growth over time.**

After the inversion of the flame front, and in concordance with the formation of distorted tulip flames, the velocity profile changes periodically. Moreover, the pressure growth accompanies the displacement velocity, i.e., a positive slope of pressure growth is associated with an increase in displacement velocity and, consequently, a negative slope indicates a decrease in velocity. At $t = 4.33\ ms$, the tulip flame is already formed and starts the formation of the DTF characterized by an increase in velocity. The formation of the first DTF takes about 1.11 ms, whereas the second one is formed in a shorter period, 0.98 ms. These results are in agreement with the experimental ones discussed in (Xiao et al., 2013). Further discussions about the relationship between pressure oscillations and velocity in closed tubes are included in (Gonzalez, 1996).

Figure 13 shows the temporal evolution of the vorticity and the pressure growth at the leading flame front. It is noticeable from this figure that at $t = 0.52\ ms$, i.e., when the flame shape starts to elongate along the x-axis, the vorticity suffers an abrupt increase. This occurs because the associated gas expansion leads to the production of vorticity. More specifically, the burnt gas behind the flame front is always rotational and depends on the radius of curvature. Additionally, the streamlines at the flame front are deviated due to the conservation of tangential momentum and the increase in the normal velocity due to mass conservation, Darrius Landau (DL) instability. The referred deviation becomes stronger when the curvature radius increases. This phenomenon is explained in detail in (Clavin & Searby, 2016). At $t = 1.30\ ms$, the first reflecting wave shocks the flame front and amplifies the vorticity. Additionally, it is created a wrinkle at the flame front, especially at the most curved region. Similar to velocity, vorticity is coupled with pressure growth at later flame stages. Hence, the formation of TF and DTF is usually associated with the increase in velocity and vorticity, and the reduction in pressure growth. The formation of wrinkles or cellular deformation is akin to Rayleigh-Taylor (RT) instabilities (Gonzalez, 1996; Xiao et al., 2013). According to Xiao et al. (2013), RT instabilities are responsible for DTF formation at later stages and depend on the aspect ratio $\alpha$ defined as the ratio between the tube length and radius. Notice as well that RT instabilities are generated by the misalignment of pressure and density gradients (Xiao et al., 2017b).

Finally, Figure 11 shows the temporal evolution of the instantaneous pressure growth at the flame front and the pressure. This figure, unlike the overall pressure growth shown in Figure 14, where a Savitzky-Golay filter (Press & Teukolsky, 1990) is applied to reduce oscillations, highlights the coupling between the flame propagation and the production of pressure waves. Initially, the pressure does not change abruptly. After the interaction of the reflecting waves with the flame front, there is an increase in the instantaneous pressure growth. Nevertheless, this pressure growth increase does not affect the propagation dynamics of the leading flame-tip. After the flame inversion, the amplitude of the pressure growth increases considerably and agrees with previous numerical findings (Xiao et al., 2015).

**4.6. Formation of tulip (TF) and distorted tulip (DTF) flames**

The formation of tulip flames involves the inversion of the flame front from a finger shape to a concave one with a cusp pointing to the right end wall (Ponizy et al., 2014). As highlighted above, there are several factors that affect the propagation dynamics of premixed flames. Figure 15 shows the evolution of the streamlines near the flame front. After the skirt flame touches the tube lateral walls, the fluid inside the burnt region moves in two opposite directions, Figure 15 ($t = 2.53\ ms$), the fluid near the flame front expands towards the fresh gas, whereas the fluid near the left end wall moves in opposite direction. As is shown in Figure 15, $t = 2.60\ ms$, this reverse flow in the burnt gas generates a single vortex near the side walls, which moves towards the flame front. The referred vortex reaches the flame front at $t = 2.90\ ms$, that is when the flame front starts to flatten. The phenomenon just described is crucial for tulip flame formation. In addition, at $t = 3.17\ ms$, the vortex under discussion generates a reverse flow in the fresh gas region. The influence of the pressure gradient in this reverse flow was observed by Kurdyunov and Matalon (2015) in their study carried out accounting for open narrow channels. The vortex formed creates a strong coupling between burnt and the fresh gases. Thus, reverse flows dominate the flame propagation dynamics at the centerline. According to Xiao(Xiao et al., 2015) indeed, reverse flows are responsible for both flame inversion and tulip flame formation.

Figure 16 shows in turn velocity distributions and streamlines during the formation of TF and DTF. It is first observed form this figure that the vortex generated by the skirt touch accompanies the flame propagation, and the magnitude of the velocity at the vortex regions changes according to the flame development stage. Thus, at $t = 4.25\ ms$, when the tulip flame is being formed, the flow velocity near to the lateral wall is around 3500 cm/s, maximum velocity in the tube, and the burnt gas pushes the flame in the lower region. At $t = 4.31$ then, the velocity at the side wall is reduced considerably, to 2000 cm/s, and, at $t = 4.39$, this flow velocity is about 750 cm/s. Consequently, in agreement with Figure 14, the tulip flame is fully formed when the displacement velocity of the leading flame tip is completely decelerated. In contrast, the flow velocity at the centerline increases when the leading flame tip decelerates and decreases when the flow velocity near the lateral wall increases its velocity. These results agree with previous studies such the one by Matalon and Metzener (2001). They suggest indeed that the TF formation is due to the large vortex formed at the early stages of the flame development fostered by the enclosure of the lateral walls.

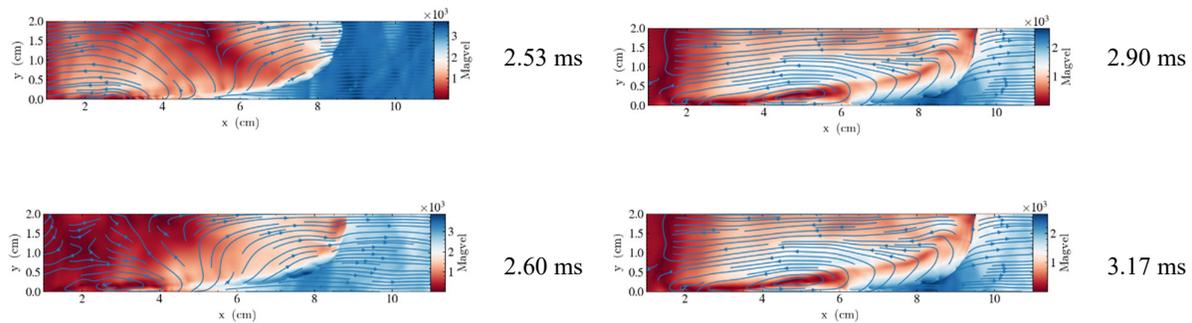

**Figure 15 Streamlines and velocity magnitude distribution during a flame inversion.**

The formation of the first DTF follows a similar mechanism to that one described in the case of TF. The interaction of the reverse flow, vortex, and flame front produces the required conditions for its formation. After the sudden deceleration of the leading flame-tip and the reverse flow velocity increase in the burnt region, it is created a strong burnt gas flux at the bottom of the channel that pushes the fresh gas toward the right end wall, Figure 16 ($t = 4.72\ ms$). As the velocity difference between both burnt and fresh gas fluxes increases, the vortex generated contributes to the growth of the first DTF. Additionally, a vortex behind the flame front that interacts with its surroundings and produces wrinkles along the flame surface is also generated, Figure 16 ($t = 4.82\ ms$). These wrinkles become more intense as the displacement velocity increases. However, it is observed that the cusp formed at the bottom wall is the one that governs the flame propagation dynamics. Finally, at $t = 4.93\ ms$, compared to the flow at the side wall, the reverse flow increases its magnitude considerably, so the displacement velocity decelerates. This mechanism repeats and forms a second cusp (Figure 14). The oscillation period of the velocity profile at later flame development stages defines the formation of DTF (Xiao et al., 2013). Notice that for the formation of the first DTF here, the oscillation period was $1.10\ ms$.

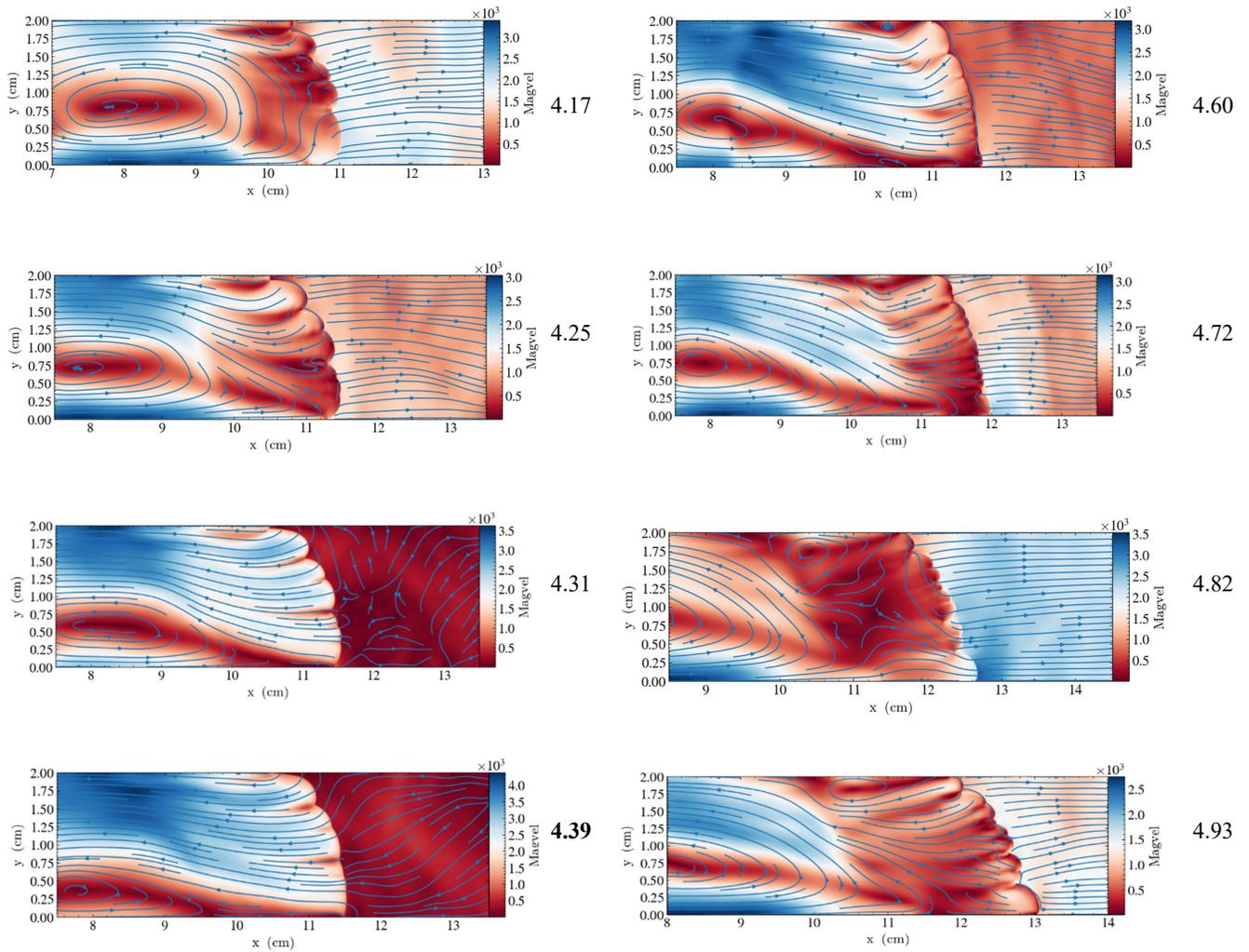

**Figure 16** Streamlines and velocity magnitude distribution near the flame front during the formation of a DTF.

## 5. CONCLUSIONS

The propagation of premixed flames in closed channels was numerically studied in this work. More specifically, to have an insight about the associated reactive flow dynamics, the formation of tulip and distorted tulip flames was analyzed. All numerical simulations were carried out with the open-source computational tool PeleC. In particular, a fully compressible reactive 2D Navier-Stokes equation system, coupled with a 21-steps chemical kinetic mechanism, was solved. The numerical simulations were performed using PPM (piecewise parabolic method) as numerical scheme and employing adaptative mesh refinement (AMR) processes. The interaction between pressure waves, vorticity, and flame fronts was particularly evaluated. In terms of results, the ones obtained here highlight a flame dynamic similar to others discussed in prior experimental and numerical studies: the initial spherical flame first evolves to an elongated one, and then to tulip and distorted tulip flames. Nevertheless, unlike other numerical results obtained in the past, where the flame front is modeled as a hydrodynamic discontinuity, thickened flame model, the ones obtained in this work show the formation of wrinkles (usually associated with DL and thermos-diffusive instabilities), especially at the initial stages of the flame development and where the flame front radius of curvature is higher. Notice that the streamlines deviation caused by the wrinkles tend to intensify the initial flame deformation. Likewise, it has been also observed that the pressure waves reflected by the channel walls interact with the flame front increasing locally the vorticity; however, the propagation dynamics of the leading flame-tip is not affected by these pressure waves.

In addition, the obtained results illustrate the formation of a relatively large-scale vortex after the flame skirt touches the channel side walls. This vortex then expands and travels towards the flame front. As a result, a reverse flow is generated at the channel centerline whose velocity magnitude increases over time, which causes the displacement velocity of the leading flame-tip to decrease sharply. This mechanism creates favorable conditions to produce a flame inversion. Furthermore, as a result of the pressure gradient, a vortex ahead of the flame front is also generated, which increases the reverse flow velocity. Thus, the reverse flow is the main responsible for the flame inversion, and naturally, the main mechanism which generates it. However, there is also another mechanism that foster the TF formation, that is, Rayleigh-Taylor instabilities. These instabilities are produced by the pressure waves created by the interaction between the skirt flame and the channel lateral walls. DTF formation initiates in turn with a strong reverse advection motion at the center of the channel due to the large vortex. Additionally, a strong vortex motion is also produced behind the flame front, which results in a temporal positive displacement in the burnt region and the formation of multiple wrinkles along the flame surface. The referred vortex is dispelled by the pressure waves reflected by the lateral walls. This formation mechanism is repeated to form a second DTF, but this time at a different formation period. Overall, the propagation dynamics of premixed flames are mainly influenced by hydrodynamics and thermos-diffusive instabilities, and by the interaction of reflected pressure waves with the curved flame front. This interaction mainly predominates in the formation of TF and DTF, since it triggers into unstable regimens, such as those characterized by the presence of Rayleigh-Taylor instabilities.

## 6. ACKNOWLEDGEMENTS

This work has been supported by the US Army Research Laboratory under Research Grant No. W911NF-22-1-0275. Luis Bravo was supported by the US Army Research Laboratory 6.1 Basic research program in propulsion sciences.

## 7. REFERENCES

Bychkov, V., Akkerman, V., Fru, G., Petchenko, A., & Eriksson, L.-E. (2007). Flame acceleration in the early stages of burning in tubes. *Combustion and Flame*, *150*(4), 263–276. https://doi.org/10.1016/j.combustflame.2007.01.004

Chung, L. (2006). *Combustion Physics*. Cambridge University Press.

Clanet, C., & Searby', G. (1996). *On the "Tulip Flame" Phenomenon*.

Clavin, P., & Searby, G. (2016). *Combustion Waves and Fronts in Flows*. Cambridge University Press. https://doi.org/10.1017/CBO9781316162453

Colella, P., & Woodward, P. R. (1984). The Piecewise Parabolic Method (PPM) for Gas-Dynamical Simulations. In *Journal of Computational Physics* (Vol. 54).

Dos, R. E., Schneider, A., Celis, C., Armando, A., & Zevallos, M. (n.d.). *ENC-2022-0104 Thermodynamic Modelling of Rotating Detonation engine Cycles*.

Ellis, O. C. de Champleur. (1900). Flame movement in gaseous explosives mixtures... . In *Flame movement in gaseous explosives mixtures...* [S.n.].

Ern, A., & Giovangigli, V. (1995). Fast and Accurate Multicomponent Transport Property Evaluation. *Journal of Computational Physics*, *120*(1), 105–116. https://doi.org/10.1006/jcph.1995.1151

Gamezo, V. N., Ogawa, T., & Oran, E. S. (2007). Numerical simulations of flame propagation and DDT in obstructed channels filled with hydrogen-air mixture. *Proceedings of the Combustion Institute*, *31 II*, 2463–2471. https://doi.org/10.1016/j.proci.2006.07.220

Gonzalez, M. (1996). *Acoustic Instability of a Premixed Flame Propagating in a Tube*.

H Markstein, B. G. (1953). The Initiation and Growth of Explosions in Liquid and Solid. In *Pron. Japan. Fifth Natl. Cong. for Appl. Mechanics* (Vol. 367, Issue 10). Physica.

Han, W., Wang, H., Kuenne, G., Hawkes, E. R., Chen, J. H., Janicka, J., & Hasse, C. (2019). Large eddy simulation/dynamic thickened flame modeling of a high Karlovitz number turbulent premixed jet flame. *Proceedings of the Combustion Institute*, *37*(2), 2555–2563. https://doi.org/10.1016/j.proci.2018.06.228

Henry de Frahan, M. T., Rood, J. S., Day, M. S., Sitaraman, H., Yellapantula, S., Perry, B. A., Grout, R. W., Almgren, A., Zhang, W., Bell, J. B., & Chen, J. H. (2022). PeleC: An adaptive mesh refinement solver for compressible reacting flows. *International Journal of High-Performance Computing Applications*. https://doi.org/10.1177/10943420221121151

Kurdyumov, V. N., & Matalon, M. (2015). Self-accelerating flames in long narrow open channels. *Proceedings of the Combustion Institute*, *35*(1), 921–928. https://doi.org/10.1016/j.proci.2014.05.082

Li, J., Zhao, Z., Kazakov, A., & Dryer, F. L. (2004). An updated comprehensive kinetic model of hydrogen combustion. *International Journal of Chemical Kinetics*, *36*(10), 566–575. https://doi.org/10.1002/kin.20026

Li, X., Xiao, H., Duan, Q., & Sun, J. (2021). Numerical study of premixed flame dynamics in a closed tube: Effect of wall boundary condition. *Proceedings of the Combustion Institute*, *38*(2), 2075–2082. https://doi.org/https://doi.org/10.1016/j.proci.2020.08.032

Marc T, & Jon S Rood. (n.d.). *PeleC: An adaptive mesh refinement solver for compressible reacting flows*. Retrieved April 10, 2023, from https://github.com/AMReX-Combustion/PeleC

Metzener, P., & Matalon, M. (2001). Premixed flames in closed cylindrical tubes. *Combustion Theory and Modelling*, *5*(3), 463–483. https://doi.org/10.1088/1364-7830/5/3/312

Oran, E. S., & Gamezo, V. N. (2007). Origins of the deflagration-to-detonation transition in gas-phase combustion. *Combustion and Flame*, *148*(1–2), 4–47. https://doi.org/10.1016/j.combustflame.2006.07.010

Poinsot, T., & Veynante, D. (n.d.). *Theoretical and Numerical Combustion Second Edition*.

Ponizy, B., Claverie, A., & Veyssière, B. (2014). Tulip flame - the mechanism of flame front inversion. *Combustion and Flame*, *161*(12), 3051–3062. https://doi.org/10.1016/j.combustflame.2014.06.001

Press, W. H., & Teukolsky, S. A. (1990). Savitzky-Golay Smoothing Filters. *Computers in Physics*, *4*(6), 669. https://doi.org/10.1063/1.4822961

van Leer, B. (1979). Towards the ultimate conservative difference scheme. V. A second-order sequel to Godunov's method. *Journal of Computational Physics*, *32*(1), 101–136. https://doi.org/10.1016/0021-9991(79)90145-1

Xiao, H., Houim, R. W., & Oran, E. S. (2015). Formation and evolution of distorted tulip flames. *Combustion and Flame*, *162*(11), 4084–4101. https://doi.org/10.1016/j.combustflame.2015.08.020

Xiao, H., Houim, R. W., & Oran, E. S. (2017a). Effects of pressure waves on the stability of flames propagating in tubes. *Proceedings of the Combustion Institute*, *36*(1), 1577–1583. https://doi.org/10.1016/j.proci.2016.06.126

Xiao, H., Houim, R. W., & Oran, E. S. (2017b). Effects of pressure waves on the stability of flames propagating in tubes. *Proceedings of the Combustion Institute*, *36*(1), 1577–1583. https://doi.org/10.1016/j.proci.2016.06.126

Xiao, H., Makarov, D., Sun, J., & Molkov, V. (2012). Experimental and numerical investigation of premixed flame propagation with distorted tulip shape in a closed duct. *Combustion and Flame*, *159*(4), 1523–1538. https://doi.org/10.1016/j.combustflame.2011.12.003

Xiao, H., Wang, Q., He, X., Sun, J., & Shen, X. (2011). Experimental study on the behaviors and shape changes of premixed hydrogen-air flames propagating in horizontal duct. *International Journal of Hydrogen Energy*, *36*(10), 6325–6336. https://doi.org/10.1016/j.ijhydene.2011.02.049

Xiao, H., wang, Q., Shen, X., Guo, S., & Sun, J. (2013). An experimental study of distorted tulip flame formation in a closed duct. *Combustion and Flame*, *160*(9), 1725–1728. https://doi.org/10.1016/j.combustflame.2013.03.011

Zhang, W., Almgren, A., Beckner, V., Bell, J., Blaschke, J., Chan, C., Day, M., Friesen, B., Gott, K., Graves, D., Katz, M., Myers, A., Nguyen, T., Nonaka, A., Rosso, M., Williams, S., & Zingale, M. (2019). AMReX: a framework for block-structured adaptive mesh refinement. *Journal of Open Source Software*, *4*(37), 1370. https://doi.org/10.21105/joss.01370

**8. RESPONSIBILITY NOTICE**

The view and conclusions contained in this document are those of the authors and should not be interpreted as representing the official policies or position, either expressed or implied, of the U.S. Army Laboratory or the U.S. Government. The U.S. Government is authorized to reproduce and distribute reprints for Government purposes not withstanding any copyright petition herein.